\documentclass[rmp,aps,twocolumn,nofootinbib,floats]{revtex4}
\usepackage{epsfig}

\begin{document}
\title{Theory of Quantum Corrals and Quantum Mirages}

\author{Gregory A. Fiete}
\email{fiete@physics.harvard.edu}
\affiliation{Department of Physics, Harvard University, Cambridge, MA 02138}
\author{Eric J. Heller}
\email{heller@physics.harvard.edu}
\affiliation{Department of Physics and Department of Chemistry and Chemical
Biology, Harvard University, Cambridge, MA 02138}

\begin{abstract}  
Quantum corrals are two dimensional structures built atom by atom on
an atomically clean metallic surface using a scanning tunneling
microscope (STM).  These two dimensional structures ``corral''
electrons in the surface states of noble metals, leading to standing
wave patterns in the electron density inside the quantum corral. We
review the physics of quantum corrals and relate the signal of the STM
to the scattering properties of substrate electrons from atomic
impurities supported on the surface. The theory includes the effects
of incoherent surface state electron scattering at the impurities and
quantitively describes nearly all of the current STM data on quantum corrals,
including the recent quantum mirage experiments with Kondo effect.  We
discuss the physics underlying the recent mirage experiments and
review some of the outstanding questions regarding the Kondo effect
from impurities in nanoscale structures on metallic surfaces.  We also
summarize recent work on variations of ``quantum'' corrals: optical
corrals and acoustical corrals.
\end{abstract}                                                                 

\maketitle
\tableofcontents

\section{Introduction}
\label{sec:intro}
Quantum corrals are the beautiful result of a marriage between
technology and basic science.  They are built atom by atom (using
approximately 30-80 atoms) on atomically smooth metallic surfaces
using a scanning tunneling microscope (STM)\footnote{The developers of
the STM, Gerd Binnig and Heinrich Rohrer, were awarded the Nobel Prize
for Physics in 1986.}.  Once the corrals are built, the
STM\footnote{For a more detailed discussion of the STM
see \textcite{book:chen}.} can be used to study these nanometer scale
structures with atomic resolution in space and better than meV
resolution in energy.  The data of the STM can be rendered in false
color to produce breathtaking images\footnote{For a stunning
demonstration of the sorts of images that can be produced with STM
data see: http://www.almaden.ibm.com/vis/stm/catalogue.html.} that
reveal standing wave patterns of coherent electrons inside the
corrals.

The history of quantum corrals begins with the pioneering work of
\textcite{eigler} who were the first to demonstrate that the STM could
be used to controllably move atoms from place to place on the surface
of a substrate.  Not long afterwards \textcite{crommie} built the
first quantum corrals from iron atoms on the Cu(111) surface and
imaged standing wave patterns inside them.  In the early experiments
it was thought that ``stadium'' shaped corrals could be used as a
laboratory to study ``quantum
chaos''~\cite{rick,rick_2,crommie_2,crommie_3} but the walls proved
too leaky (and the states of the corrals too low in energy) for the
electrons to bounce around the (unstable) periodic orbits long enough
to detect any ``scarring'' effects~\cite{heller_2}.  A very intriguing
recent STM corral experiment was done by \textcite{hari} who
combined the physics of quantum corrals with the Kondo effect to
achieve a beautiful ``mirage'' inside the corral of the spatially
localized spectroscopic response of a Kondo impurity where there was
in fact no Kondo impurity.  The mirage experiment achieves this by
taking advantage of both the locally modified electron density in the
corral\footnote{\textcite{kliewer} studied the effect of the
corral-modulated surface state electron density on the spectroscopy of
Mn on Ag(111), which did not display a Kondo effect.} and the
scattering properties of a Kondo impurity.

In this Colloquium we review the scattering theory of STM measurements of
quantum corrals including the recent mirage experiments with Kondo effect.
We demonstrate the success of the scattering theory in reproducing {\em every}
 detail of the 
experiments including the electron standing wave patterns, the energies and 
widths of corral states and all features of the quantum mirage.  The 
scattering theory we present is based on a single-particle picture but takes
the many-body physics of the Kondo effect into account phenomenologically 
in a straightforward way. At the end of this colloquium we 
discuss extensions of the quantum corrals to optical corrals 
and acoustical corrals.  We begin our discussion with a
review of the important physics of the substrate on which 
quantum corrals are built.

\section{The Importance of Surface States}
\label{sec:surface}
The beautiful standing wave patterns observed in STM corral
experiments~\cite{crommie,rick,hari,kliewer} result from the presence
of Shockley surface states\footnote{For more details
see~\textcite{book:davison} and for experimental results for several
materials see~\textcite{kevan}.  The surface states themselves are
still a very active area of research with many STM studies being
reported in recent years: \textcite{li_surf,buergi_surf,kliewer_surf}.}
on the metallic substrate. These are the same surface states
responsible for the standing wave patterns observed near a step
edge~\cite{avouris}. Surface states are the result of a particular
crystallographic cut of the material, usually a noble metal, which
places the Fermi energy in a band gap for electrons propagating normal
to the surface.  The surface states of Cu(111), Au(111) and Ag(111)
are commonly used in STM experiments.  In the direction normal to the
surface (and in a range of angles around the normal), Bloch states are
forbidden at the Fermi energy. However, solutions to the Schr\"odinger
equation exist with exponentially decaying amplitude into both the
bulk material and the vacuum. For such solutions electrons are still
free to move in the plane of the surface and form a type of two
dimensional electron gas (2DEG) there.  Often, the surface state band
is only partially filled, giving a low density on the surface, and a
nearly quadratic dispersion relation with a constant effective mass.

The scattering theory that we develop for quantum corrals in
Sec.~\ref{sec:scatt} is based on these free two dimensional surface
state electrons.  We will see that although the quantum corrals are
two dimensional systems in many respects, there are some important ways in
which the underlying bulk material makes its presence felt.  This is
especially true with the quantum mirage experiments where the bulk
electrons play an important role in the formation of the Kondo
resonance~\cite{knorr}.

Before we leave our brief discussion of surface states it is important
to give some typical numerical values of important quantities such as
the wavelength of electrons in the surface states, $\lambda$, the
effective mass of surface state electrons, $m^*$, and density of
states of the surface state electrons, $\varrho_{\rm surf}$.  These
three quantities are all related through the dispersion relation
\begin{equation}
E_{\rm surf}(k)-E_F=E_0 + \frac{\hbar^2 k^2}{2 m^*}\;,
\label{eq:dispersion}
\end{equation}  
where $\varrho_{\rm surf}=\frac{m^*}{\pi \hbar^2}$ (for $E>E_0$)
includes both spin up and spin down electrons. In the case of Cu(111),
Au(111) and Ag(111) the surface state band minimum, $E_0$, is very
close to the Fermi energy.  Typical values are fractions of an eV
below the Fermi energy~\cite{kevan}, $E_F$, where $E_F$ is measured
relative to the bottom of the bulk state bands and is typically 5-10
eV.  For Cu(111), $E_0 \approx -450$ meV and $m^*=0.38m_e$ with $m_e$
the mass of the free electron. The surface state electron density of
Cu(111) is $n \approx7 \times 10^{13}\, {\rm cm}^{-2}$ which
corresponds to approximately one surface state electron per 12 \AA\
$\times$ 12 \AA\ square.

There are three important physical consequences of small $E_0$.  The
first is that it makes the dispersion relation quadratic and {\em
isotropic} (in the plane of the surface) to a very good approximation.
An isotropic dispersion relation is very convenient for the
application of scattering theory because one does not need to know the
orientation of the underlying crystal lattice.  Secondly, a small
$E_0$ makes the filling of the surface state band rather low compared
to bulk bands, which in turn makes the typical wavelength of the
surface state electrons, $\lambda \approx \lambda_F= \frac{2\pi}{k_F}$,
very large compared the lattice spacing and the size of atomic
impurities on the surface.  For Cu(111), $\lambda_F=29.5$~\AA.  Since
$\lambda_F$ is much larger than the underlying Cu(111) lattice
spacing, the standing wave patterns are easy to separate from atomic
scale charge density variations and since $\lambda_F$ is large
compared to the surface adatoms, we can make an s-wave approximation
in the scattering theory.\footnote{The scattering theory we describe
below is a two dimensional theory.  The dynamics in the direction
normal to the surface is assumed to be energetically inaccessible, much
like the case of a 2 dimensional electron gas (2DEG) that forms at the
interface of GaAs/AlGaAs. Electron scattering out of the plane of the
surface (into the bulk) is taken into account in a phenomenological
way in the scattering theory by adding an imaginary component to the
phase shift.  This is discussed in detail in Sec.~\ref{sec:scatt}.}
Thirdly, a small $E_0$ makes the electron filling small so the density
of surface states is small compared to bulk states at the same energy.
This has implications for the microscopic details of the Kondo effect
from a magnetic impurity like Co on the Cu(111) surface.  We will
return to this point in Sec.~\ref{sec:mirage_theory}.  We now turn to
the STM measurement.

\section{STM Theory: Topographic Images and Spectroscopic Measurement}
\label{sec:STM}
\begin{figure}[t]
\begin{center}
\epsfxsize=8.5cm
\epsfbox{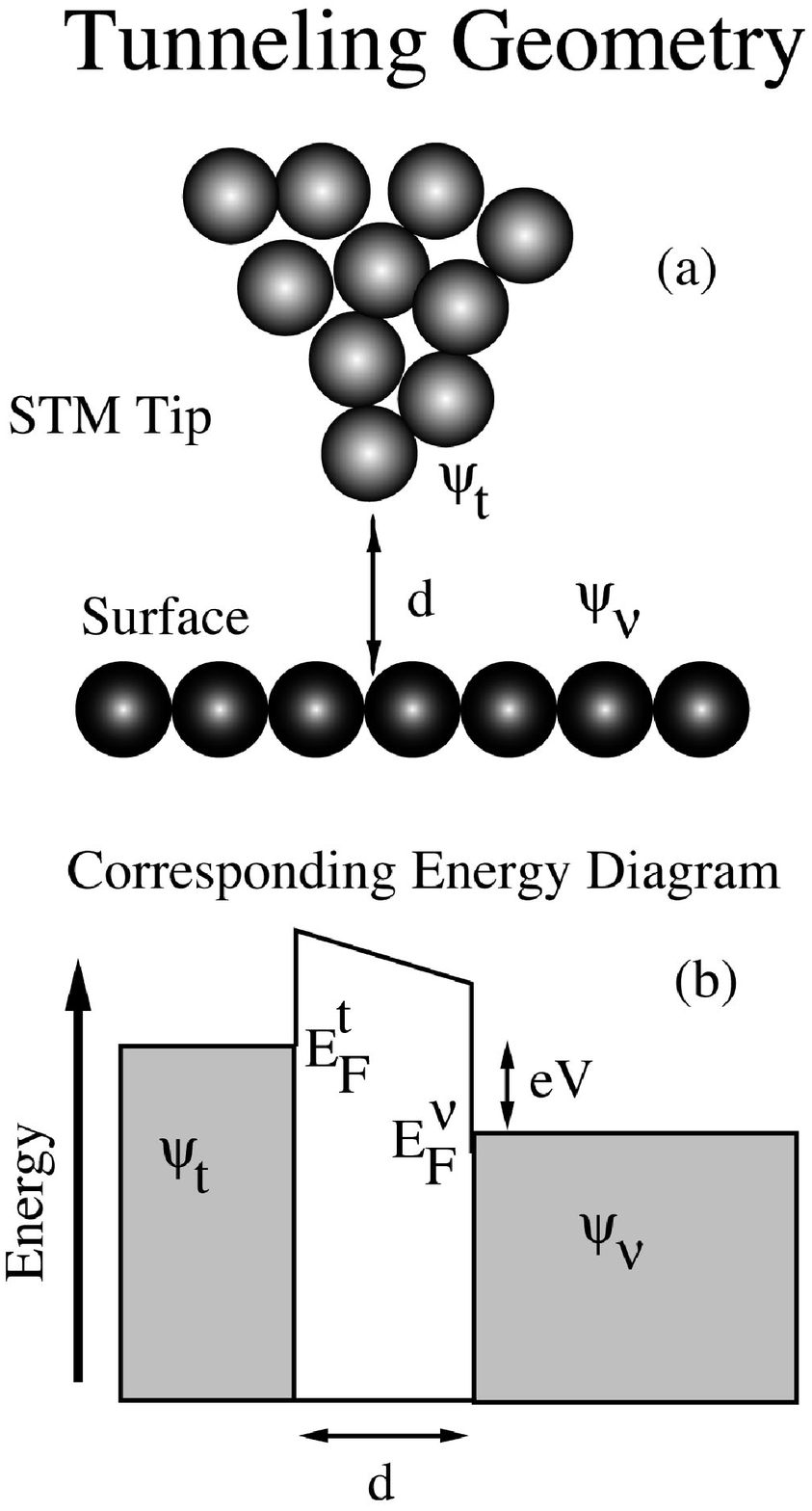}
\end{center}
\caption{Geometry of the scanning tunneling microscope
measurement and energy diagram. (a) Schematic of the STM tip above the
substrate.  The STM tip states are labeled by $\psi_t$ and the
eigenstates of the substrate are labeled by $\psi_\nu$.  The current
is exponentially sensitive to the tip-surface distance, $d$. (b)
Energy diagram of the tunneling process.  Electrons must tunnel across
a vacuum barrier of thickness $d$ from occupied states of the tip to
unoccupied states of the surface (energies $E_F^\nu <\epsilon <
E_F^t$).  The total current, Eq.~(\ref{eq:current}), is given by the
sum of all such processes, while the conductance,
Eq.~(\ref{eq:conductance}), just measures the tunneling rate for
electrons at a particular energy in this window.
\label{stm}}
\end{figure}

In this section we briefly review the physics of the tunneling measurement.
The basic tunneling geometry and energy diagram is shown in Fig.~\ref{stm}.
The STM tip usually sits a few~\AA\ above the surface.  The STM data
can be taken in two ways: (i) A feedback loop can be used to control
the height of the tip above the surface so that the total tunneling
current is kept constant as the tip is scanned over the surface.  This
is called a ``topographic'' image and, as we will soon see, at each
point it is a measure of the {\em energy-integrated} local density of
surface states.  (ii) In the second type of measurement the feedback
loop is opened so that the tip height is kept roughly constant with
respect to the surface and the voltage is swept to measure the local
spectroscopy at the tip position.

Tunneling measurements of quantum corrals are typically done at small
voltage biases, $V < 0.3$ Volts, and low temperatures, $T < 70$ K.  In
such a situation perturbation theory can be applied to compute the
tunneling current in terms of the unperturbed tip states and surface
states. According to Fermi's Golden Rule, the current at position
${\bf r}$ and STM bias voltage $V$ is~\cite{tersoff,kleber}
\begin{equation}
I({\bf r})={ 2 \pi e \over \hbar} \sum_{t,\nu} |M_{t,\nu} ({\bf r})|^2 
f(\epsilon_t) \left (1-f(\epsilon_\nu)\right ) \delta(\epsilon_t +eV- \epsilon_\nu)\;,
\label{eq:current_general}
\end{equation}
where $e$ is the charge of the electron, $t$ ($\nu$) labels the tip
(surface) states, $f$ is the Fermi function and $M_{t,\nu} ({\bf r})$
is the matrix element from the tip state $t$ to the surface state
$\nu$ at position ${\bf r}$.  The expression,
Eq.~(\ref{eq:current_general}), has a simple physical interpretation.
It says that the tunneling current is proportional to the square of the
matrix element connecting the various tip states to the various
surface states times a factor which gives the probability of an
occupied tip state and an empty surface state.  The delta function
enforces energy conservation. Finally all tip states and surface
states are summed over. When the tip is treated as a point source,
$|M_{t,\nu} ({\bf r})|^2 \propto |\psi_\nu({\bf r})|^2$
\cite{tersoff}, where $\psi_\nu({\bf r})$ are the eigenfunctions of
the surface.  Assuming also that the temperature is low enough to
replace the Fermi functions by step functions, using the relation
$\int d\omega \delta(\epsilon_t +eV-\omega)\delta(\omega
-\epsilon_\nu)=\delta(\epsilon_t +eV-\epsilon_\nu)$ and converting the
sum over tip states to an integral, we obtain
\begin{equation}
I({\bf r}) \propto \int_0^{eV} \varrho_t(\epsilon) {\rm LDOS}({\bf r}, \epsilon) d\epsilon \;,
\end{equation}
where $\varrho_t(\epsilon)$ is the density of states of the tip and
the local density of states (LDOS) is given by
\begin{equation}
{\rm LDOS}({\bf r}, \epsilon)=\sum_\nu |\psi_\nu({\bf r})|^2\delta(\epsilon - E_\nu)\;.
\label{eq:ldos_stm}
\end{equation}
Usually the density of states of the tip is assumed constant so it can
be pulled out of the integral,
\begin{equation}
I({\bf r}) \propto \int_0^{eV}{\rm LDOS}({\bf r}, \epsilon) d\epsilon\;,
\label{eq:current}
\end{equation}
 and
\begin{equation}
{dI \over dV}({\bf r},\epsilon) \propto {\rm LDOS}({\bf r}, \epsilon)\;.
\label{eq:conductance}
\end{equation}

The last three equations above,
Eqs.~(\ref{eq:ldos_stm}),~(\ref{eq:current}),
and~(\ref{eq:conductance}), are the most important formulas for the
interpretation of the quantum corral experiments. The central quantity
to calculate is Eq.~(\ref{eq:ldos_stm}) as the current,
Eq.~(\ref{eq:current}), and the conductance,
Eq.~(\ref{eq:conductance}), depend on it.  The LDOS is expressed in
terms of the eigenstates, labeled by $\nu$, of the surface.  It is
through the calculation of these eigenstates from scattering theory
that Eq.~(\ref{eq:ldos_stm}) provides the bridge between scattering
theory and the tunneling measurement of the STM.  We will develop this
connection fully in Sec.~\ref{sec:scatt}.  From Eq.~(\ref{eq:current})
and Eq.~(\ref{eq:conductance}) it is evident that the STM signal is
related to the square of the surface state wavefunctions at a given
location.  If the wavefunction has large (small) amplitude at a
particular location the current and conductance will tend to be larger
(smaller) there.

A topographic measurement corresponds to Eq.~(\ref{eq:current}) in
which a feedback loop is used to modulate the tip height to keep the
current constant.  This is usually the type of measurement used to
produce data like the standing wave patterns in quantum corrals.
Typical bias voltages are on the order of 10 meV so that the current
at position ${\bf r}$ is an integral over approximately 10 meV of
energy. In most experiments, the density of states at any given
position ${\bf r}$ does not vary much over 10 meV.  However, in the
mirage experiments the Kondo effect actually produces strong
variations in the local density of states over 10 meV~\cite{hari}.

In the spectroscopic measurement the STM tip-surface distance is held
fixed by turning off the feedback loop. The voltage is swept (at a
given position) to reveal the energy dependence of the LDOS,
Eq.~(\ref{eq:conductance}).  This is the type of measurement that
reveals the energies and widths of resonances in quantum corrals which
appear as peaks in a plot of dI/dV vs. V.  The Kondo resonance at a
Kondo impurity also has a strong signature in
dI/dV~\cite{li,madhavan,hari}. The quantum mirage is most easily
probed in this way~\cite{hari,fiete}.

\section{Scattering Theory for Surface State Electron Density}
\label{sec:scatt}
In this section we develop a scattering theory for the electron
density in quantum corrals.  In Sec.~\ref{sec:surface} we emphasized
the importance of the surfaces states on the (111) surfaces of noble
metals and gave the important properties for the development of
scattering theory: two dimensional electron states on the surface,
isotropic and parabolic dispersion of the energy and long electron
wavelength compared to the lattice spacing and the size of the
adatoms.  We now describe how the quantities of the STM measurement
given in Sec.~\ref{sec:STM}, namely Eq.~(\ref{eq:ldos_stm}), are
obtained from scattering theory.  The physical picture to have in mind
is of a circularly symmetric electron amplitude emanating from the STM
tip into the surface states of the substrate.\footnote{By measuring
the change in the differential conductance when the voltage is swept
from below surface state band energies to above the lowest surface
state band energy, the relative fraction of flux into the surface
states and bulk states can be determined.  It is typically around 50\%
\cite{knorr,buergi_fabry}.}  This amplitude spreads radially outward
from the tip until it encounters a defect (such as an impurity) on the
surface or a step edge, at which time it scatters.  Part of this
amplitude is reflected back to the STM tip\footnote{The exponential
decay of the surface states into the vacuum can affect the details of
the topographic measurements due to the feedback
loop~\cite{kliewer_njp}.} (possibly scattering several more times
along the way from different impurities) and interferes with the
outgoing amplitude leading to fluctuations in the LDOS, and hence the
tunneling current, as a function of position.  Note that the
fluctuations are a result of the {\em coherent} part of the
back-scattered amplitude.

Let the Hamiltonian of an electron on the surface be $\hat H=\hat
H_0+\hat V$, where $\hat H_0$ is the Hamiltonian describing free
propagation in the surface states and $\hat V$ accounts for the
spatially local and separate potential perturbations due to the
impurities on the surface.  The amplitude to propagate from point
${\bf r}$ to point ${\bf r'}$ in time $t$ on the surface is given by
the retarded Green's function, $G^{\rm ret}({\bf r'}, {\bf r},t)=-i
\theta(t) \langle {\bf r'}|e^{-i \hat H t/\hbar}|{\bf r}\rangle$, where
$\theta(t)$ is the step function.  The eigenstates of $\hat H$ are
the scattering eigenstates of the particle in the presence of the
potential $\hat V$.  Inserting a complete set of eigenstates,
\begin{equation}
G^{\rm ret}({\bf r'}, {\bf r},t)=-i \theta(t) \sum_\nu \langle {\bf r'}|
e^{-i  E_\nu t/\hbar}|\psi_\nu\rangle \langle \psi_\nu|{\bf r}\rangle\;,
\end{equation}
and taking the Fourier transform of this,
\begin{equation}
G^{\rm ret}({\bf r'}, {\bf r},\epsilon)=\sum_\nu {\psi^*_\nu({\bf r}) \psi_\nu({\bf r'}) \over \epsilon - E_\nu + i\delta}\;.
\label{eq:greens}
\end{equation}
Here $\psi_\nu({\bf r})$ are the eigenstates of the Hamiltonian $\hat
H$ and $\delta$ is an infinitesimal positive quantity.  For the STM
measurements, we are interested in the part of the amplitude that back
scatters to the tip.  Thus, we are interested in ${\bf r'}= {\bf r}$.
The imaginary part of the diagonal amplitude is proportional to the
local density of states,
\begin{equation}
{\rm LDOS}({\bf r},\epsilon )\equiv -{1\over \pi} {\rm Im}\left [G^{\rm ret}({\bf r}, {\bf r},\epsilon)\right ]
=\sum_\nu |\psi_\nu({\bf r})|^2\delta(\epsilon  - E_\nu)\;.
\label{eq:ldos}
\end{equation}

What we have established is a relationship between the full Green's
function, Eq.~(\ref{eq:greens}), the scattering eigenstates of the
Hamiltonian $\hat H$, $\psi_\nu({\bf r})$, and the local density of
states.  What remains is to develop a method for calculating the
Green's function, Eq.~(\ref{eq:greens}).

We first consider the case where $\hat V$ represents a single scatterer.  
Dyson's equation can be written
\begin{equation}
\hat G^{\rm ret}=\hat G^{\rm ret}_0 +\hat G^{\rm ret}_0 \hat V \hat G^{\rm ret}\;,
\label{eq:lippmann_op}
\end{equation}
where $\hat G^{\rm ret}$ is the full retarded Green's function and
$\hat G^{\rm ret}_0$ is the free retarded Green's function.  When
$\hat V$=0, $\hat G^{\rm ret}=\hat G^{\rm ret}_0$.  The $\hat G^{\rm
ret}$ on the right hand side of Eq.~(\ref{eq:lippmann_op}) can be
formally eliminated by iterating the equation.  In operator
notation,
\begin{eqnarray}
\hat G^{\rm ret}&=&\hat G^{\rm ret}_0 + \hat G^{\rm ret}_0 \hat V \hat G^{\rm ret}_0 + \hat G^{\rm ret}_0 \hat V \hat G^{\rm ret}_0 \hat V \hat G^{\rm ret}_0
+ \cdots \\ \nonumber
&=& \hat G^{\rm ret}_0 +\hat G^{\rm ret}_0 (\hat V + \hat V \hat G^{\rm ret}_0 \hat V + \cdots)\hat G^{\rm ret}_0.
\end{eqnarray}
The terms in the series have the physical interpretation of a particle
that (i) does not scatter at all from the potential, (ii) scatters
once and leaves, (iii) scatters once, propagates, scatters again and
then leaves, (iv...) and so on to infinite order. Truncation of the
series at $\hat V$, for example, is just the first Born approximation.
The terms within parentheses can be grouped into into a single object
called the {\em t-matrix}. The {\em t-matrix} is defined by
\begin{equation}
\hat T=\hat V + \hat V \hat G^{\rm ret}_0 \hat V + \cdots\;.
\end{equation}
When the spatial extent of the scattering potential is small compared
to the wavelength of the incoming particle, as is the case for adatoms
on the Cu(111) surface, the scattering is s-wave (isotropic) because
the wavelength of the incident particle is too large to ``feel'' the
spatial structure of the target.  In the s-wave approximation, the
{\em t-matrix} takes a particularly simple form in position
representation~\cite{thaler}:
\begin{eqnarray}
G^{\rm ret}({\bf r},{\bf r}) &=& G_0^{\rm ret}({\bf r},{\bf r}) + \nonumber \\
&&  \int \int d^2{\bf r'}
d^2{\bf r''}G_0^{\rm ret}({\bf r},{\bf r'})\nonumber \\
&& \times s \delta({\bf r_0}-{\bf r''}) \delta({\bf r_0}-{\bf r
  '})G_0^{\rm ret}({\bf r''},{\bf r})\;,
\label{eq:t_matrix}
\end{eqnarray}
where $s(k)=\frac{4i\hbar^2}{m^*}(e^{2i\delta(\epsilon)}-1)$, ${\bf
r_0}$ is the position of the impurity and $\delta(\epsilon)$ is the
energy dependent phase shift ($\epsilon(k)$ is given by
Eq.~(\ref{eq:dispersion})) in the s-wave orbital channel (which can be
computed once $V(r)$ is known or determined directly from experiment).
The integral can then be done trivially to yield
\begin{equation}
G^{\rm ret}({\bf r},{\bf r}) = G_0^{\rm ret}({\bf r},{\bf r}) + s G_0^{\rm ret}({\bf r},{\bf r_0}) 
G_0^{\rm ret}({\bf r_0},{\bf r}).
\end{equation}
Note that when $V(r)$ goes to zero, $\delta(\epsilon)$ goes to zero
and one obtains $G^{\rm ret}({\bf r}, {\bf r}) = G_0^{\rm ret}({\bf
r},{\bf r})$. That is, the full Green's function reduces to the free
Green's function.

\begin{figure}[h]
\begin{center}
\epsfxsize=8cm
\epsfbox{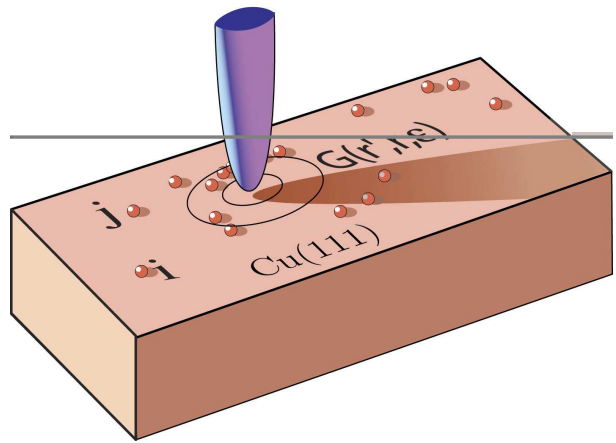}
\end{center}
\caption{
Schematic of the scattering geometry of
multiple scattering theory.  The scattering centers, shown as little
beads, are adatoms on the surface of a noble metal such as Cu(111).
In the approximation that the STM tip is point-like, a circularly
symmetric electron amplitude, $G_0({\bf r'}, {\bf r}, \epsilon)$,
emanates from the tip into the surface states of the metal and
encounters the impurities on the surface.  Since the wavelength of the
electrons in the surface states is much larger than the size of the
scatters, one can treat the scatterers as s-wave scatterers and ignore
all higher orbital channels. Because the scatterers are far apart
compared to their size, we assume that electrons propagate freely
between impurities. $i$ and $j$ label different impurities.
\label{scatt}}
\end{figure}

The extension to several scatterers is straightforward. The schematic 
situation is shown in Fig.~\ref{scatt}.
The new ingredient in the many scatterer case is an extra
self-consistency condition on the scattered amplitude. Imposing this
self-consistency condition is equivalent to calculating the scattering
among all the impurities to infinite order. This is the heart of
multiple scattering theory. (The {\em t-matrix} gives the result of
scattering from a single impurity to infinite order.)

In the presence of $N$ scatterers the {\em t-matrix} equation
$\sum_{i=1}^N \hat T_i \hat G^{\rm ret}_i= \sum_{i=1}^N \hat V_i \hat
G^{\rm ret}$ ($\hat V_i$ are non-overlapping scattering potentials, the
$\hat T_i$ are the corresponding {\em t-matricies} for these
potentials) generalizes Eq.~(\ref{eq:t_matrix}) to
\begin{eqnarray}
G^{\rm ret}({\bf r},{\bf r}) &=& G_0^{\rm ret}({\bf r},{\bf r}) + \sum_{i=1}^N \int \int d^2{\bf r'}
d^2{\bf r''}G_0^{\rm ret}({\bf r},{\bf r'})
\nonumber \\
&& \times s_i \delta({\bf r}_i-{\bf r''}) \delta({\bf r}_i-{\bf r '})G^{\rm ret}_i({\bf r''},{\bf r}) \nonumber\\
     &  = &G_0^{\rm ret}({\bf r},{\bf r}) + \sum_{i=1}^N s_i G_0^{\rm ret}({\bf r},{\bf
  r}_i)G^{\rm ret}_i({\bf r}_i,{\bf r})\;,
\label{eq:G_final}
\end{eqnarray}
where the $G^{\rm ret}_i$ are the self-consistently calculated values of the 
Green's functions at the locations of the scatterers,
\begin{equation}
G^{\rm ret}_i({\bf r}_i,{\bf r})=G_0^{\rm ret}({\bf r}_i,{\bf r}) + \sum_{j\neq i}^N s_iG_0^{\rm ret}({\bf r}_i,{\bf
         r}_j)G^{\rm ret}_j({\bf r}_j,{\bf r})\;,
\label{eq:Gi}
\end{equation}
and 
\begin{equation}
s_i(\epsilon)=\frac{4i\hbar^2}{m^*}(e^{2i\delta_i(\epsilon)}-1)\;,
\label{eq:phase}
\end{equation}
for the $i^{th}$  scatterer. The solution of Eq.~(\ref{eq:Gi}) is given
by the equation
\begin{equation}
{\bf G=A^{-1}G_0}\;,
\label{eq:inverse}
\end{equation}
where ${\bf A}$ is an $N \times N$ matrix with elements ${\bf
A}_{ij}=\delta_{ij} -s_iG_0({\bf r}_i,{\bf r}_j)$ containing all the
information about the propagation between the impurities and ${\bf
G_0}$ and ${\bf G}$ are $N$-dimensional column vectors of elements
${\bf G_{0i}}=G_0({\bf r}_i,{\bf r})$ and ${\bf G}_i=G({\bf r}_i,{\bf
r})$, respectively, containing the information about propagation from
the STM tip to the impurities and from the impurities to the STM tip.

The STM signal is then calculated from scattering theory by specifying
the s-wave scattering phase shifts $\delta_i(\epsilon)$, the locations
$\{ {\bf r}_i\}$ of the $N$ impurities on the surface and the incident
electron amplitude.  Given the dispersion relation,
Eq.~(\ref{eq:dispersion}), the free Green's function, $G_0^{\rm
ret}({\bf r'},{\bf r},\epsilon)$, is determined from
Eq.~(\ref{eq:greens}) in the case of $\hat V$=0.  In two dimensions,
the outgoing Green's function from a point source is $G_0^{\rm
ret}({\bf r'},{\bf r},\epsilon)=-i\frac{m^*}{2
\hbar^2}\left(J_0(k|{\bf r'}-{\bf r}|)+i Y_0(k|{\bf r'}-{\bf r}|)
\right)$ where $J_0$ ($Y_0$) is the Bessel function of the first
(second) kind.  The final step is to fix the energy and then solve the
system of equations, Eq.~(\ref{eq:G_final}) and Eq.~(\ref{eq:Gi}) by
Eq.~(\ref{eq:inverse}) at the particular chosen energy. (All three
of these equations depend on the energy and must be re-solved for each
new energy.)  The solution is then substituted into
Eq.~(\ref{eq:ldos}), which directly gives the STM signal through
Eqs.~(\ref{eq:current}) and~(\ref{eq:conductance}).

The theory just developed applies equally well to electrons or holes
near the Fermi energy.  Although the STM tip is the source (or sink in
the case of positive bias, i.e., the tip has larger voltage) of
electrons (or holes), we have not included one correction that in
principle is present, namely any residual unscreened potential felt by
an electron near the STM tip.  In fact the tip itself can be thought
of as a source of scattering, causing disturbances to any electron
passing under it. However, we have so far not seen any experimental
evidence indicating this correction is needed at small bias
voltages.\footnote{At larger biases, there may be some Stark shifting
of the states due to the electric field of the STM tip: There are
differences in the surface state band edge energies for STM and
photoemission experiments.}

\section{Application to Quantum Corrals}
\label{sec:corrals}
The scattering theory of Sec.~\ref{sec:scatt} may be directly applied
to quantum corrals.  Here we discuss the case of Fe atoms on Cu(111)
\cite{rick} which do not show a Kondo effect at 4 K. Our goal is to
calculate the standing wave patterns and corral spectroscopy of the
type first observed by~\textcite{crommie}.  To do so we pull together
the results of Secs.~\ref{sec:surface},~\ref{sec:STM}
and~\ref{sec:scatt}.  Since we know the dispersion relation,
Eq.~(\ref{eq:dispersion}), for the Cu(111) surface as well as the
positions of the iron impurities from STM measurements, all that
remains to determine the current and conductance at a given position
is a determination of the phase shift, $\delta(\epsilon)$, of the Fe
atoms.  Once $\delta(\epsilon)$ is determined, the ${\rm LDOS}({\bf
r}, \epsilon)$ is determined everywhere by the scattering theory
except within 7~\AA\ of an adatom, where there is extra charge density
not accounted for in the theory.  Electron amplitude from the STM tip
is assumed to emanate in a circularly symmetric fashion into the
surface states, so we use the outgoing free Green's function,
$G_0^{\rm ret}({\bf r'},{\bf r},\epsilon)=-i\frac{m^*}{2
\hbar^2}\left(J_0(k|{\bf r'}-{\bf r}|)+i Y_0(k|{\bf r'}-{\bf r}|)
\right)$, as the incident amplitude.

\begin{figure}[th]
\begin{center}
\epsfxsize9cm
\epsfbox{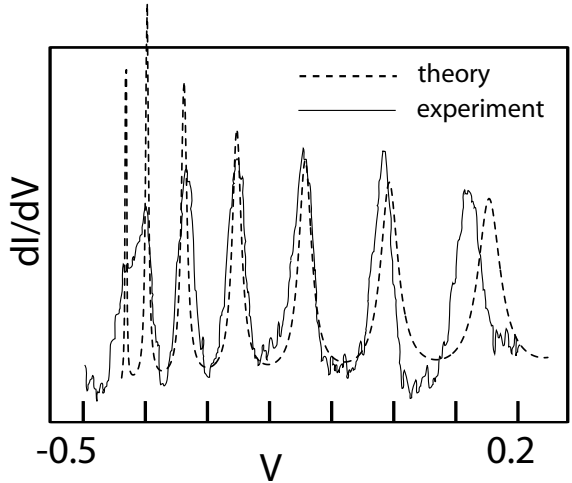}
\end{center}
\caption{
The experimental and theoretical voltage
dependence of dI/dV, with the tip of the STM located at the center of
an 88.7~\AA\ diameter, 60-atom circular corral of Fe atoms on a
Cu(111) surface. A smooth background has been removed from the
experimental data. 
\label{spectrum}}
\end{figure}

\begin{figure}[th]
\begin{center}
\epsfxsize7cm
\epsfbox{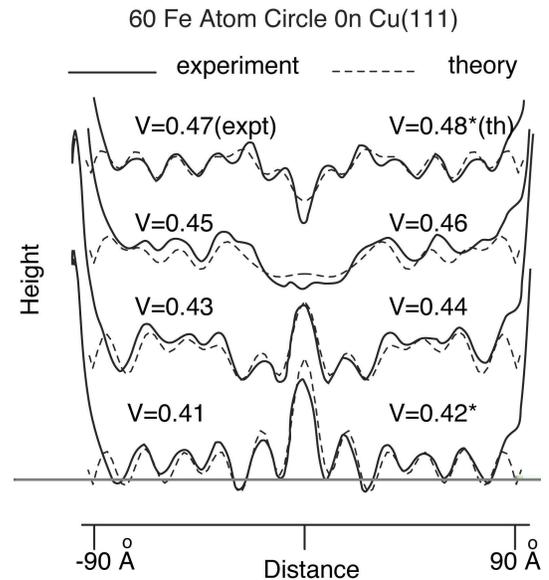}
\end{center}
\caption{\label{cross_section} The experimental data and theoretical curves
for the tip height as a function of position across the diameter of a
circlular corral (88.7~\AA\ diameter, 60-atoms) for low bias voltages.
Various voltages are given; they are measured relative to the bottom
of the surface state band.  The asterisk (*) on the first and last theory
voltage values is to call attention to the slight shift relative to
experiment used to obtain the best fits.}
\end{figure}

\begin{figure}[t]
\begin{center}
\epsfxsize8cm
\epsfbox{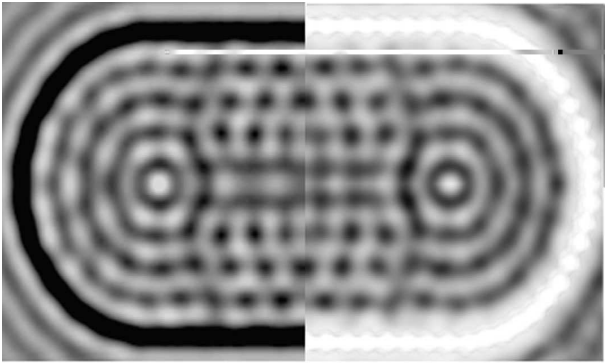}
\end{center}
\caption{\label{stadium} Local density of electron states (LDOS) near
$E_F$ for a 76 Fe atom ``stadium'' of dimensions 141 $\times$
285~\AA. Right Hand Side: Experiment, bias voltage 0.01 V
($\epsilon$=0.45 eV); Left Hand Side: Theory ($\epsilon$=0.46 eV). The
density at the locations of the Fe adatoms is not accounted for in the
theory and appears black.}
\end{figure}

Early measurements~\cite{eigler_2} of single iron impurities on the
surface of Cu(111) pointed to a phase shift near -$80^o$.  However,
from Eq.~(\ref{eq:phase}), it is clear the the Green's function is
invariant with respect to a phase shift of $\pi$, so the phase shift
could equally well have been near +100$^o$.  When the scattering
theory was applied with a phase shift of +100$^o$ to circular corrals
to compute dI/dV, Eq.~(\ref{eq:conductance}), the widths of the
resonances were far too narrow compared to experiment, indicating a
longer electron confinement than was actually inferred from the
broader, measured linewidths.  The important insight~\cite{rick} was
that the resonances could be broadened if one allows electron
amplitude to be absorbed from the surface states at the Fe impurities.
A phase shift of nearly +100$^o$ is quite close to +90$^o$.  This
leads to
\begin{equation}
  s_i(\epsilon)=\frac{4i\hbar^2}{m^*}(e^{2i\delta_i(\epsilon)}-1) 
\stackrel{\delta= \frac{\pi}{2}}{ \longrightarrow} \frac{4i\hbar^2}{m^*}(-2)\;.
\end{equation}
On the other hand, if the Fe atoms were assumed maximally absorbing 
``black dots''~\cite{rick}, $\delta=i\infty$, 
\begin{equation}
s_i(\epsilon)=\frac{4i\hbar^2}{m^*}(e^{2i\delta_i(\epsilon)}-1)
\stackrel{\delta= i\infty}{ \longrightarrow} \frac{4i\hbar^2}{m^*}(-1)\;,
\end{equation}
so that the overall scattering amplitude has the same phase but is
reduced by a factor of 2.  Thus, the two phase shifts, $\delta=
\frac{\pi}{2}$ and $\delta=i\infty$, are equivalent except that the
``black dot'' approximation, $\delta=i\infty$, leads to an attenuation
of the scattered wave and a broadening of corral resonance widths.
When $\delta=i\infty$ is used to evaluate the ${\rm LDOS}({\bf r},
\epsilon)$, at the center of a circular quantum corral the agreement
with experiment is excellent.  Fig.~\ref{spectrum} shows a direct
comparison between theoretical and experimental dI/dV curves for an
88.7~\AA\ diameter, 60-atom circular corral of Fe atoms on a Cu(111)
surface.  Note that except for the first peak\footnote{The first peak
has been investigated in more detail by \textcite{crampin_2}.} the
agreement with experiment is excellent.  Both the resonance energies
and the widths of the resonances are remarkably alike and scale
together except for the highest energy peak.\footnote{The scattering
theory can be brought into nearly perfect agreement with even the
highest energy peak by allowing for a quartic correction to the
parabolic dispersion, Eq.~(\ref{eq:dispersion})~\cite{stella}.}
Fig.~\ref{cross_section}
shows a comparison between theory and experiment for a ``topographic''
image for a cut across the diameter of the same circular corral.  Note
again the excellent agreement: Every experimental oscillation is
quantitatively reproduced by the scattering theory. Finally, the full
standing wave patterns for both theory and experiment for a
``stadium'' shaped quantum corral are shown in Fig.~\ref{stadium}.

\begin{figure}[h]
\begin{center}
\epsfxsize8cm
\epsfbox{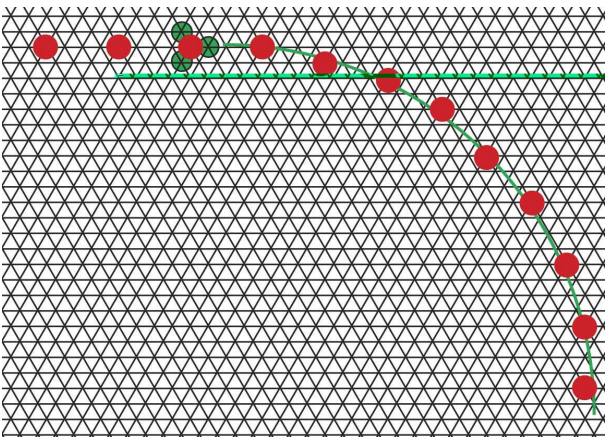}
\end{center}
\caption{\label{grid} Shown is a grid of the underlying lattice of the
Cu(111) surface.  Adatoms cannot sit at exact positions of an ideal
stadium, ellipse or circular shaped corral, but must sit on the
nearest site of the underlying lattice.  When the exact positioning of
the adatoms in taken into account in the theoretical calculations, the
agreement with experiment is enhanced.  At one site, 4 dark circles
are shown.  The lighter circles represent possible positions for the
darker, central adatom.  Obviously the central position is best for
the geometry given by the solid line shown.}
\end{figure}

The Fe adatoms can be located only at the available triangular lattice
sites in the Cu(111) surface.  This lattice allows arcs, ellipses, and
other shapes to be only approximated. The locations where one can
place atoms can be seen in Fig.~\ref{grid},
for the case of a 48 atom stadium, where the smooth boundary is drawn
for comparison.  It is important to put in the correct atomic
positions in order to get the best agreement with the experiments.
The corral walls, while acting like smooth (although absorbing)
boundaries for some purposes, still reveal their roughness and
granularity.

We now turn to a physical interpretation of the ``black dot''
approximation.  If electron flux is absorbed at the Fe atoms where
does it go?  We believe that much of the lost surface state amplitude
goes into to the bulk.\footnote{ Spin-flip processes at the Fe
impurities would also appear as a loss of coherent amplitude.}  This
idea has been supported by theoretical studies of~\textcite{crampin}
and~\textcite{hormandinger}. Shortly after the work
of~\textcite{rick}, ~\textcite{harbury} developed an elastic
scattering theory of quantum corrals.  The elastic theory is able to
qualitatively reproduce the standing wave patterns inside the corrals
but does relatively poorly compared to the inelastic scattering theory
for dI/dV.  (See Fig. 3 in~\textcite{harbury}.)  The inelastic
scattering theory presented here accounts well (at energies higher
than the first one or two peaks) for the widths and heights of the
resonances in corrals compared to the elastic theory
of~\textcite{harbury}.  However, at lower energies there is
disagreement due to the intrinsic lifetime of the surface states that
saturates the linewidths~\cite{crampin_2}.  This effect can be
exploited in quantum corrals and near step edges to study many-body
effects in the surface states~\cite{crampin_2}. We will return to this
point in Sec.~\ref{sec:recent}.

\begin{figure}
\begin{center}
\epsfxsize=7cm
\epsfbox{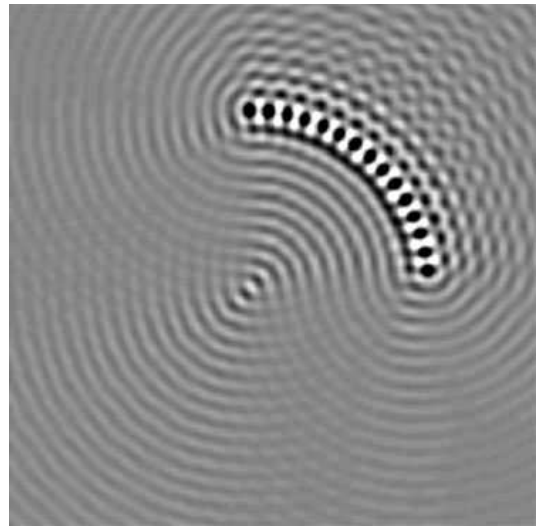}
\end{center}
\caption{\label{sparse_atoms} A theoretical calculation of STM data that
could not be modeled with a ``particle-in-the-box'' approximation
(because it is ``open'') or any type of smooth arc (note the circular
``wavelets'' near the impurities).}
\end{figure}

It is important to summarize what we have learned from the application
of scattering theory to quantum corrals thus far: (i) Corrals do
confine electrons in surface states, but do so rather poorly
(resonance widths are broad) because the adatoms tend to couple
surface states quite strongly to bulk states.  A host of studies
\cite{crampin,buergi_fabry,hormandinger,fiete,knorr,schnieder} suggest
that it is quite generic for adatoms to strongly couple surface states
to bulk states. (ii) The standing wave patterns in corrals depend on
coherent electron propagation in the surface states to give
interference effects.  For temperatures below 70 K, the coherence
length of surface state electrons on noble metals is hundreds of
Angstroms \cite{jeandupeux}, while the corrals typically have maximum
dimensions on the order of a hundred Angstroms thus allowing coherent
electron propagation across the corrals.  (iii) ``Particle-in-a-box''
models \cite{crommie} may qualitatively agree with the observed
resonance energies of closed structures, but they have no predictive
power for resonance widths or standing wave patterns in open
structures.  For example, consider the arc in Fig.~\ref{sparse_atoms};
 it shows predicted STM data with a variety of features that certainly
could not be modeled with a box, or even, for some of the features,
with a smooth arc imposing some boundary condition.  Scattering theory
works equally well for one atom as for any arbitrary arrangement of
any number of atoms (provided the structure is small enough to allow
coherent electron propagation across it). (iv) The only place the
scattering theory fails to agree with experiment is within 7~\AA\ of
an atom.  Here the assumptions of the theory break down because the
extra charge density at the impurity is not properly accounted for.

One comment is in order on the multiple scattering theory.  As simple
as it is to invert a matrix of the dimension equaling the total
number of atoms to obtain the Green's function, it is still perhaps
curious to do a multiple scattering expansion in terms of $0, 1, 2,
\dots$ scattering events.  It turns out that this fails for typical
configurations, due to the presence of closely spaced pairs, triplets,
etc., of atoms.  Even though the Q-factor of the cavity, as indicated
for example by the line widths of the dI/dV resonances, is only around
$2$, suggesting about two bounces are important before leakage
occurs
%\footnote{The quality factor is defined as frequency of a
%resonance divided by 2 $\pi$ times the power lost per cycle. In
%quantum mechanics, this is just the frequency attempted..{\bf Rick
%needs to fill this in}}
, the Q-factor and the low order of scattering
is an appropriate concept only for walls thought of as smooth
scattering units, with the local multiple internal scattering between
neighboring atoms included to infinite order.

%If one performs a calculation for a plane wave of appropriate
%wavevector impinging normal to a line of Fe adatoms at the closest
%experimental Fe atom spacing, one finds that only about 25\%
%of the amplitude is coherently backscatted (also normal to the line),
%while 25\% is transmitted, and 50\% is lost to decohering inelastic
%events.

The scattering theory does not have to confine itself only to atomic
surface impurities.  The experiments abound with step
edges~\cite{buergi_fabry,jeandupeux,burgi,morganstern}, for example,
even though one looks for regions as far away as possible from such
defects to build corrals. The step edges affect the images, although
not so much inside closed corrals, which have enough attenuation at
the wall to prevent those paths which begin inside the corral, get
out, hit an edge, and come back inside from having any important
weight.

\section{The Mirage Experiment}
\label{sec:mirage_experiment}
A recent and interesting variation of the original quantum corral
experiments were the ``quantum mirage'' experiments\footnote{For a
more detailed description of the experiment we recommend that readers
consult the original paper: ~\textcite{hari}.}
of~\textcite{hari}. The quantum mirage experiments make use of the low
temperature physics associated with a magnetic ion (e.g., Fe, Co, Mn)
in electrical contact with a bulk metal (e.g., Cu, Au, Ag): the so
called Kondo Effect. In the quantum mirage experiments,
~\textcite{hari} built an elliptical corral with magnetic atoms (Co)
which exhibit a Kondo effect at 4 K on Cu(111).

The Kondo effect is the many-body response of the free electrons in
the Fermi sea to the magnetic impurity; It is intimately
related to spin-flip scattering events of free conduction electrons
from the magnetic ion.  To understand the problem in detail takes a
substantial investment of time, but fortunately the results of a
detailed analysis relevant to the quantum mirage can be stated quite
simply and succinctly: (i) The spin of the conduction electrons tend
to become anti-correlated (oppositely aligned) with the spin of the
magnetic impurity so that at low temperatures (when the Kondo effect
is present) the local spin of the magnetic ion is fully or at least
partially screened.  An important special case is when the spin of the
ion is 1/2.  Then, the Kondo effect {\em completely} screens it at
sufficiently large distances.\footnote{Larger spins can also be
completely screened, but require that more than 1 orbital channel of
the conduction electrons to couple to them.  See \textcite{blandin}
for a discussion.}  In the scattering approach that we are using, this
means that spin-flip scattering is ``frozen out'' and we can treat the
scattering as purely {\em potential scattering} (i.e. we neglect
spin-flip scattering processes).  We will discuss in
Sec.~\ref{sec:kondo} how this ``freezing out'' of the spin might be
understood. (ii) The impurity density of states (the density of states
of the atomic d- or f-levels that give rise to the magnetic moment)
develops a narrow resonance near the Fermi energy that is often termed
the ``Kondo Resonance''.  This resonance is picked up in the STM
measurement and is the main spectroscopic signature of Kondo atoms.

The narrow Kondo resonance (whose width is related to the Kondo energy
scale, $T_K\sim 50$ K for Co on Cu(111)) appears in dI/dV near a Kondo
atom.  It had been observed near (within 10~\AA) isolated atoms
\cite{li,madhavan} earlier, but~\textcite{hari} used the modification
of the surface state electron density in an elliptical quantum corrals
to produce a spectroscopic ``mirage'' inside the corral of a Kondo
atom where there was in fact no Kondo atom (the ``source'' of the
mirage was a Co atom inside the corral more than 70~\AA\
away).\footnote{The experimental and theoretical spectroscopic
signature is shown in Fig.~\ref{onatom}.}  In order to fully understand the
mirage experiment, we must first review some details and essential
results of the theory of the Kondo effect.  We will need these results
for the application of our scattering theory to adatoms which show a
Kondo effect.

\section{Essentials of Kondo Physics}
\label{sec:kondo}
\subsection{The Anderson Model}
The Kondo effect\footnote{For an overview of Kondo effect and a list
of references see~\textcite{hewson}.  For a brief survey of the Kondo effect
in {\em mesoscopics} see~\textcite{zawadowski} and references therein.} is
the name given to the low energy response of the Fermi sea of a metal
to a magnetic impurity.  In the mirage experiments, the magnetic
impurity (Co) sat on the surface of Cu(111).  The canonical (and
simplest) model\footnote{The Anderson model applies to a spin $S=1/2$
impurity.  However, it can be shown~\cite{ujsaghy} that impurities
of higher spin can be treated with an effective spin $S=1/2$ model.
For a discussion of the Kondo effect for spin $S>1/2$
see~\textcite{blandin}.} of a local magnetic moment in a metallic host was
given by~\textcite{anderson},
\begin{eqnarray}
\hat H_{\rm Anderson} = \sum_{k,\sigma}\epsilon_{k\sigma} \hat n_{k\sigma}&& + \sum_\sigma 
\epsilon_{d}  \hat n_{d\sigma} + U  \hat n_{d\uparrow}
 \hat  n_{d\downarrow}\nonumber \\
&& +\sum_{k,\sigma}(V  \hat c^\dagger_{k\sigma} \hat d_\sigma + h.c.)\;.
\label{eq:anderson}
\end{eqnarray} 
The first term represents the energy of the electrons of the Fermi sea
(assumed to be non-interacting), the second term represents the energy
of a single localized site (an approximation to the d or f atomic
level of an atom), the third term represents an on-site repulsion if
two electrons try to occupy the localized level, and the last term
represents hybridization between the local moment and the conduction
electrons.  Here $\epsilon_{k\sigma}$ ($\hat n_{k\sigma}$) is the
energy (number operator) of an electron of the Fermi sea with
wavevector $k$ and spin $\sigma$ and $\epsilon_{d}$ ($\hat
n_{d\sigma}$) is the spin degenerate energy (number operator, not
generally spin degenerate) of an electron in the localized d or
f-level with spin $\sigma$.  Here $U$ represents the charging energy
of doubly occupying the localized level. In the fourth term, $V$ is
the hopping matrix element connecting the electrons of the Fermi sea
to the localized impurity level and vice versa, and $\hat
c^\dagger_{k\sigma}$ ($\hat d^\dagger_\sigma$) is the creation
operator for an electron in the state with wavevector $k$ (d or
f-level) with spin $\sigma$.

\begin{figure}
\begin{center}
\epsfxsize=7cm
\epsfbox{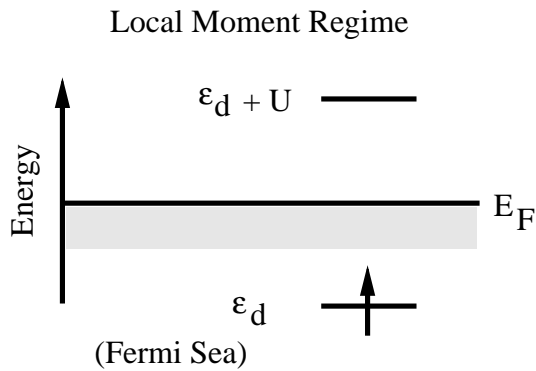}
\end{center}
\caption{\label{level}Local Moment Regime of the Anderson Model.  The Kondo
effect occurs when the impurity level energies are situated as shown.
The spin degenerate singly occupied level has energy $\epsilon_d <E_F$.
The cost for adding the second electron of opposite spin to the impurity level
is $\epsilon_d +U>E_F$.  Thus, the impurity ground state has only one 
electron on the local level giving it a net spin.}
\end{figure}

In the case $V=0$, Eq.~(\ref{eq:anderson}) can be solved exactly. The
states are just direct products of the local moment states and the
Fermi sea. The energy is just the sum of the energy of the Fermi sea
and the energy of the electron(s) on the localized level.  The energy
cost for having one electron on the localized level is $\epsilon_{d}$
and the cost for adding the second is $\epsilon_{d}+U$.  If one
considers a small nonzero $V$, the Hamiltonian is no longer exactly
solvable. The singly and doubly occupied states of the local level
will be broadened (by an amount that can be estimated by Fermi's
Golden Rule, $\Gamma \approx 2 \pi V^2 \varrho_0$, where $\varrho_0$
is the density of states of the Fermi sea at the Fermi energy, or more
precisely, as the energy $\epsilon_d$, if the density of states varies
with energy).  For the Kondo problem one particular regime of
Eq.~(\ref{eq:anderson}) is of central importance: the case where
$\epsilon_{d} < E_F$ and $\epsilon_{d}+U >E_F$. This is shown in
Fig.~\ref{level}.
~\textcite{anderson} showed that Eq.~(\ref{eq:anderson}) will lead to
local moment formation at low enough temperatures when $\Gamma \approx
2 \pi V^2 \varrho_0 \ll |\epsilon_{d}|,\, \epsilon_{d}+U $.

\begin{figure}
\begin{center}
\epsfxsize=8cm
\epsfbox{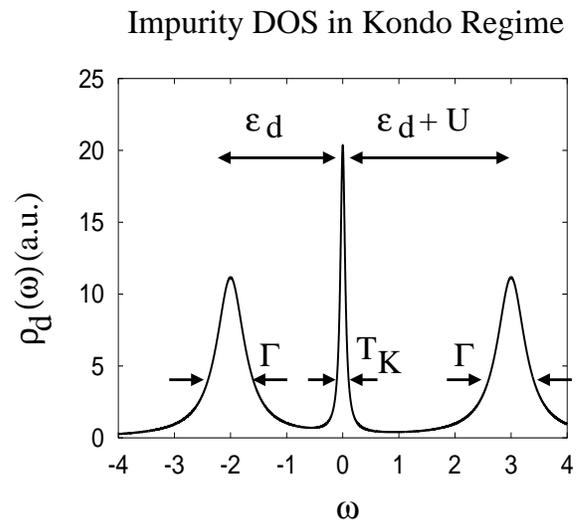}
\end{center}
\caption{
Density of States of the Anderson Impurity
Model in the Kondo Regime.The figure is not an actual calculation but
illustrates the central features of the density of states in the Kondo
regime. Both axes are in arbitrary units, but for a real system energy
units of eV would not be unrealistic.  The broad upper and lower peaks
(at energies $\epsilon_d$ and $\epsilon_d+U$) have width $\Gamma
\approx 2 \pi V^2 \varrho_0$.  These peaks are due to single-particle
energy levels of the impurity.  They are broadened by coupling to the
Fermi sea.  The central peak is a many-body resonance sometimes called
the ``Kondo peak.'' It arises from correlations beyond a mean-field
calculation such as Hartree-Fock and its width is exponentially small
in the coupling parameter $J\varrho_0$, $T_K \sim D e^{-{1\over
2 J\varrho_0}}$.
\label{dos_1}}
\end{figure}

The impurity (d-level) density of states in the Anderson model in the 
local moment regime we have just discussed is shown in Fig.~\ref{dos_1}.
The peak in the density of states at zero bias is sometimes referred
to as the ``Kondo'' peak~\cite{hewson}.  The Kondo peak always sits
near the Fermi energy and corresponds to the formation of the {\em
many-body} Kondo state.  It is this peak that shows up in the form of
a ``Fano resonance'' in the dI/dV spectra near (within 10~\AA) a Kondo
atom on the surface of a
metal~\cite{li,madhavan,hari,ujsaghy,schiller,plihal}.  Although the
Kondo resonance is associated with many-body correlations of the Fermi
sea and has no single-particle level analogous to the two spectral
peaks corresponding to the bare levels at $\epsilon_{d}$ and
$\epsilon_{d}+U $ in Fig.~\ref{dos_1}, it still {\em behaves} as a
single-particle resonance when it is fully formed at $T \ll
T_K$~\cite{nozieres,ujsaghy,plihal}.  It is this single-particle like
behavior or ``local Fermi liquid theory''~\cite{nozieres} of the Kondo
resonance that allows us to use a single-particle scattering theory
for the mirage experiments.  The density of states of Fig.~\ref{dos_1}
translates into a strongly energy dependent phase shift for electrons
of the Fermi sea near the Fermi energy~\cite{hewson}. For a single
impurity in a host (or on the surface) the density of states is (to a
good approximation) just the sum of the two:
$\varrho(\epsilon)=\varrho_0(\epsilon) +\varrho_{\rm imp}(\epsilon)$.
Therefore, the change in the density of states due to the impurity is
$\Delta \varrho(\epsilon)\equiv
\varrho(\epsilon)-\varrho_0(\epsilon)=\varrho_{\rm imp}(\epsilon)$.
It can be shown to equal~\cite{hewson}
\begin{equation}
\varrho_{\rm imp}(\epsilon) = {1\over \pi} {\partial \delta(\epsilon)\over \partial \epsilon}\;.
\label{eq:d_phase}
\end{equation}
The resonance at the Fermi energy in Fig.~\ref{dos_1}, for $\varrho_{\rm
imp}(\epsilon)$, can be approximated as a Lorentzian of width $\Gamma$
centered near the Fermi energy ($\epsilon_0$) and leads to a ${\rm
tan}^{-1}({\epsilon-\epsilon_0 \over \Gamma/2})$ in the phase shift
$\delta(\epsilon)$ of Eq.~(\ref{phase}) when Eq.~(\ref{eq:d_phase}) is
integrated over energy.  It is highly non-trivial that one can treat a
many-body problem like the Kondo effect phenomenologically with a
single-particle theory and a resonant phase shift.  It is the single
most important reason for the success of our approach to the quantum
mirage.

\subsection{The Kondo Model}
The Kondo model is a special limit of the Anderson model,
Eq.~(\ref{eq:anderson}), valid in the local moment regime shown in
Fig.~\ref{level}.  It was used by J.~\textcite{kondo} (hence the name)
to explain the minimum in the resistivity (as a function of
temperature) of metals with magnetic impurities. The Kondo model can
be derived by second order perturbation theory in $V$ from the
Anderson model.\footnote{ This result was first derived by
\cite{schrieffer}.}  The Kondo Hamiltonian (including a purely
potential scattering term that also appears in second order
perturbation theory) is
\begin{eqnarray}
\hat H_{\rm Kondo} = \sum_{k,\sigma}\epsilon_{k\sigma} \hat n_{k\sigma} + J \sum_{k, k'}
&&\hat S \cdot  \hat c^\dagger_{k\sigma} {\vec \tau_{\sigma\sigma'}\over 2}
\hat c_{k'\sigma'}\nonumber \\
&& + K\sum_{k, k',\sigma }\hat c^\dagger_{k\sigma}\hat c_{k'\sigma} \;,
\label{eq:kondo}
\end{eqnarray}
where the first term is the same as in Eq.~(\ref{eq:anderson}),
\begin{eqnarray}
J \approx V^2\Biggl({1\over U+\epsilon_d}-{1\over \epsilon_d}\Biggr)>0\;,
\end{eqnarray}
and
\begin{eqnarray}
K \approx -{V^2 \over 2}\Biggl({1\over U+\epsilon_d}+{1\over \epsilon_d}\Biggr)\;.
\end{eqnarray}
Here $\hat S$ is the spin operator of the impurity and $\vec \tau$ are
the Pauli spin matrices.  The crucial feature of Eq.~(\ref{eq:kondo})
is that it leads to spin-flip scattering events\footnote{These
spin-flip scattering events can also be looked at from the point of
view of the Anderson model.  A spin-flip would occur if, e.g., the
initial electron on the local level were spin up, a second spin down
electron hopped on in the intermediate state and then finally the
original spin up electron hopped off, leaving behind the spin down
electron on the local level.}  through terms like $S_x\tau_x +
S_y\tau_y=(S_+\tau_- + S_-\tau_+)/2$. These terms turn out to be
related to the apparent low temperature divergence of the resistivity
(as a function of temperature) in some metals with a low concentration
of magnetic impurities (which are able to flip the spins of
electrons). Kondo's explanation of the divergence comes by looking at
the effect of the second term of Eq.~(\ref{eq:kondo}) in a
second-order perturbative calculation of the {\it t-matrix}.  It turns
out that because $S_+S_- \neq S_-S_+$ one of the sums over the
intermediate states of the Fermi sea is cut off at the Fermi surface
leading to a logarithmic divergence in the
resistivity~\cite{kondo,hewson}.

Besides the features of the Kondo problem we have already mentioned,
one more result is worthy of note.  In order to understand the low
energy behavior of many physical systems it is often useful to
integrate out the high energy fluctuations and compensate for this by
``renormalizing'' the parameters of an effective low energy theory.
This can be quite complicated in general, but for the Kondo
Hamiltonian a particularly simple version known as ``poor man's
scaling,'' introduced by~\textcite{anderson_2}, can be used to
identify the low energy properties~\cite{hewson}.  The idea is to look
again at the second order contributions to the {\it t-matrix} from the
second term of Eq.~(\ref{eq:kondo}).  The sum over the intermediate
states of the conduction electrons contains electrons that are at the
band edges.  Anderson suggested removing a few states at the band
edges and adjusting $J$ so that the scattering amplitude remains
invariant (ignoring the potential scattering terms).  When this is
done a set of ``scaling equations'' is generated for $J$ which can
then be solved.  It turns out that a ``scaling invariant'' appears and
it is generally denoted by $T_K$ and referred to as the Kondo
temperature:
\begin{eqnarray}
T_K = D e^{-{1\over 2 J\varrho_0}}\;.
\label{eq:kondotemp}
\end{eqnarray}
The quantity $T_K$ is invariant under a rescaling of $J$ in response
to a shrinking of the bandwidth, $D$. As $D \to 0$, $J \to \infty$,
which from the second term of Eq.~(\ref{eq:kondo}) implies that the
spin-flip processes are ``frozen out'' in the low energy theory and
the scattering becomes purely potential scattering.  As before,
$\varrho_0$ is the density of states of the host at the Fermi
energy. The Kondo effect, in this simplest of models, is thus
characterized by only one energy scale, $T_K$.  This is the width of
the ``Kondo peak'' that appears in the low temperature density of
states of the Anderson model in Fig.~\ref{dos_1}, the width of the
Fano resonance~\cite{li,madhavan,hari,kawasaka,ujsaghy,schiller,plihal} and
it is also the width of the scattering resonance~\cite{fiete}, from
Eq.~(\ref{eq:d_phase}), of the Co atoms (for experimental temperatures
lower than $T_K$ when the Kondo resonance is well formed) on the
surface of Cu(111).

\section{Theory of Quantum Mirages}
\label{sec:mirage_theory}
\begin{figure}[t]
\epsfxsize=8cm
\epsfbox{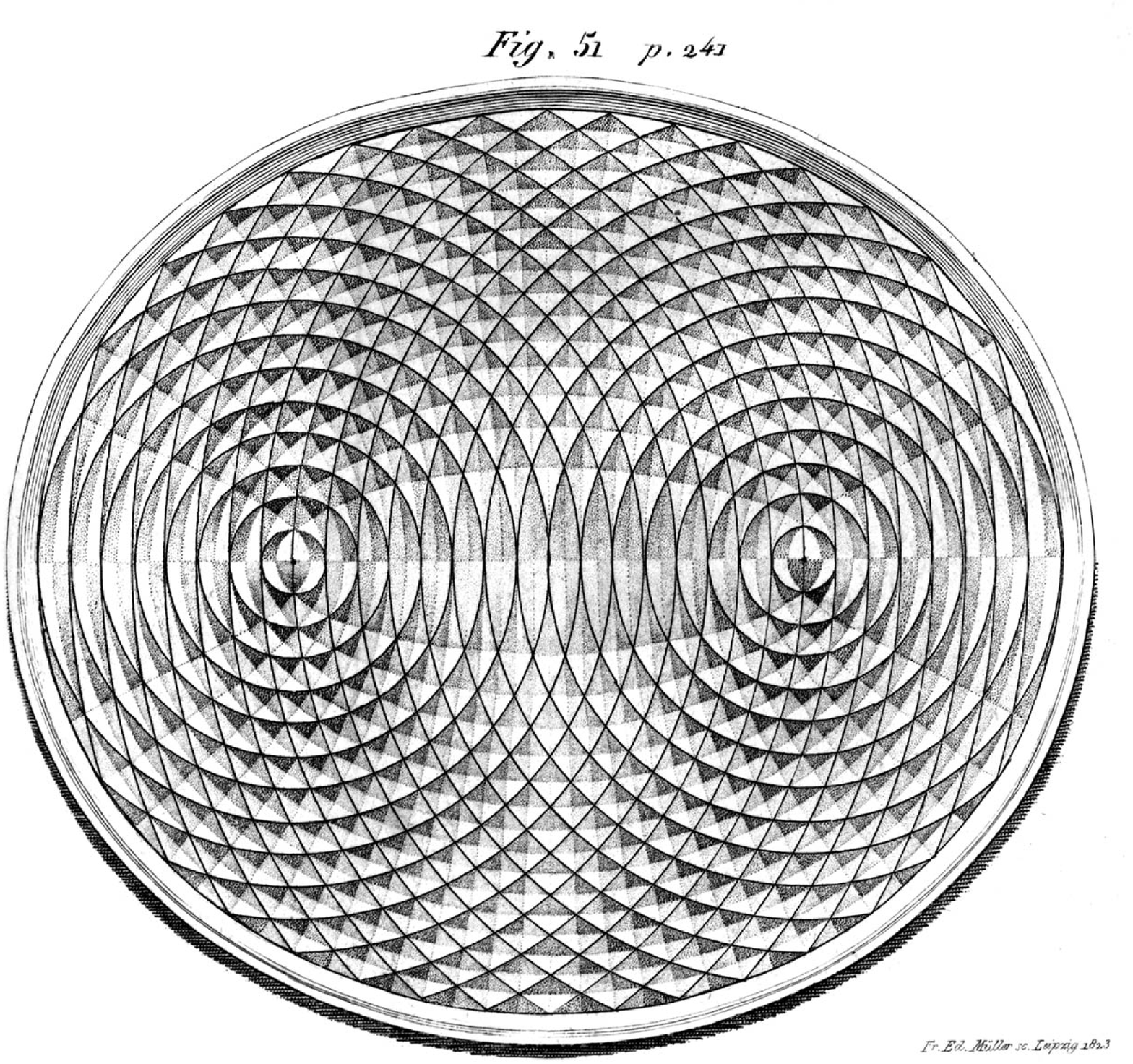}
\caption{\label{ellipse}  
A sketch from ``Wellenlehre'' (``Wave
Teachings''), an 1825 book published in Leipzig on wave theory by two
of the three Weber brothers scientists from Saxony, Ernst and Wilhelm,
showing the wave pattern of mercury waves when small amounts of
mercury are dropped in at one focus.  Notice how the other, opposite
focus looks identical, indicating that from the point-of-view of the
wave, the two foci are excited equally.}
\end{figure}

In the fascinating quantum mirage experiment \textcite{hari} decided
to use the unique scattering properties of an ellipse in an attempt to
project the properties of an atom sitting at one focus of the ellipse
to the corresponding second (empty) focus of the ellipse.  The ellipse
has been recognized for its properties in the context of waves for a
long time.  For example, the remarkable image showing surface waves of
mercury in an elliptical container (Fig.~\ref{ellipse}),
drawn by the two of the three scientifically inclined Weber brothers
in 1825 (including Wilhelm Weber, well known to physicists in
connection with electromagnetism), clearly shows the special nature of
the ellipse. This experiment almost perfectly anticipates the Eigler
group's measurements of matter waves 175 years later, since the image
corresponds to drops of mercury landing at one focus with the other
focus ``empty.''

Our theory of the quantum mirage~\cite{fiete} is based on a fairly
straightforward modification of the scattering theory originally
presented by~\textcite{rick} (for non-Kondo atoms) to account for the
Kondo effect.  As we emphasized at the end of the first subsection of
Sec.~\ref{sec:kondo}, for experimental temperatures below $T_K$ we are
able to take advantage of Nozieres' (1974) ``local Fermi liquid''
picture to write down a phenomenological single-particle theory with an
energy dependent phase shift.  Our theory of the quantum mirage
involves the following approximations, assumptions and limitations:
(i) The scattering of electrons from the adatoms is determined by a
single parameter, the s-wave phase shift, and this must be determined
from experiment or otherwise. (ii) The internal degrees of freedom
(spin) of the Kondo adatoms are ``frozen out'' at the temperature of
the experiment ($\sim 4$K) so we may use the results
of~\textcite{nozieres} to treat the Kondo atom as a potential scatterer
with a phase shift. (iii) The adatoms are far enough apart so that we
may treat the electron propagation between them as free and that RKKY
interactions are sufficiently weak that the single-impurity Kondo
physics is not altered. (iv) The theory does not include any
non-equilibrium effects and does not treat the charge density within
7~\AA\ of an atom correctly.

To make a direct comparison with experiment, we must obtain the phase
shift of the Kondo adatoms.  We do not have an {\it ab initio}
calculation of the phase shift of a single Co adatom. Rather, we fit
the resonant form of the phase shift, including inelasticity due to
the coupling of the surface states to bulk states, and calculated the
multiple scattering problem with this single atom data.

Since the on-atom electron orbital density is not accounted for in
scattering theory, we used an on-atom fit (from experimental data of a
single, isolated Co atom on Cu(111) at 4 K) involving only a
renormalization of the free-space Green's function, $G_0^{\rm
ret}({\bf r'},{\bf r},\epsilon)$, and a change in the background phase
shift to compute the STM signal on top of a Kondo adatom
\cite{kawasaka,plihal,schiller}.  This on-atom fit is not part of
our theory, but only a means of setting a reference point between
on-atom density not accounted for in our theory and the electron
density anywhere more than 7 \AA\ away from an atom on the surface
which {\it is} accounted for properly in our theory.  This fit in no
way compromises our fundamental result that the mirage is due to
resonantly scattering electrons from the Kondo atoms of the walls and
focus.  It is used only as a method of determining as accurately as
possible the phase shift of the Co on Cu(111).  Determining the phase
shift this way from experimental data constitutes a measurement of the
single Kondo atom phase shift.  We find a good fit to the s-wave phase
shift to be
\begin{equation} 
\delta(\epsilon)=\delta_{bg}+i\delta''+{\rm tan}^{-1}({\epsilon-\epsilon_0
\over \Gamma/2})\;, 
\label{phase}
\end{equation}  
where $\delta_{bg}={\pi\over 4}\pm{\pi\over10}$, $\delta''={3\over
2}\pm{1\over4}$, $\Gamma=(9 \pm 1)$ meV and $\epsilon_0$ = $E_F$ -1
meV are determined by experiment; $\delta_{bg}$ is a background phase
shift (possibly due to static charge screening at the impurity) that
controls the resonant line shape of the adatom scattering
cross-section\footnote{Recently~\textcite{schnieder} have determined
the phase shift of Co on Ag(111) which has a $T_K$ of 92 K and found
similar values to ours determined for Co on Cu(111).} and  $\delta''$ is
a measure of the inelasticity in adatom scattering and controls the
attenuation of the mirage at the empty focus.
Tan$^{-1}({\epsilon-\epsilon_0\over \Gamma/2})$ reflects resonant
scattering due to the presence of Kondo physics and can be seen to
follow directly from Eq.~(\ref{eq:d_phase}) and the density of states
shown in Fig.~\ref{dos_1}.  A similar phase shift (without the
inelastic piece, $\delta''$) would result from the model
of~\textcite{ujsaghy}.  It is likely that both bulk and surface states are
participating in the Kondo effect at an adatom\footnote{The most
recent experiments by~\textcite{knorr}, suggest that the Kondo effect at
Kondo impurities on surfaces is in fact dominated by bulk states.},
but the STM signal is more sensitive to the surface state Kondo effect
in the regime of validity of our theory ($>$ 7 \AA\ away from adatom).

\begin{figure}[t]
\epsfxsize=8cm
\epsfbox{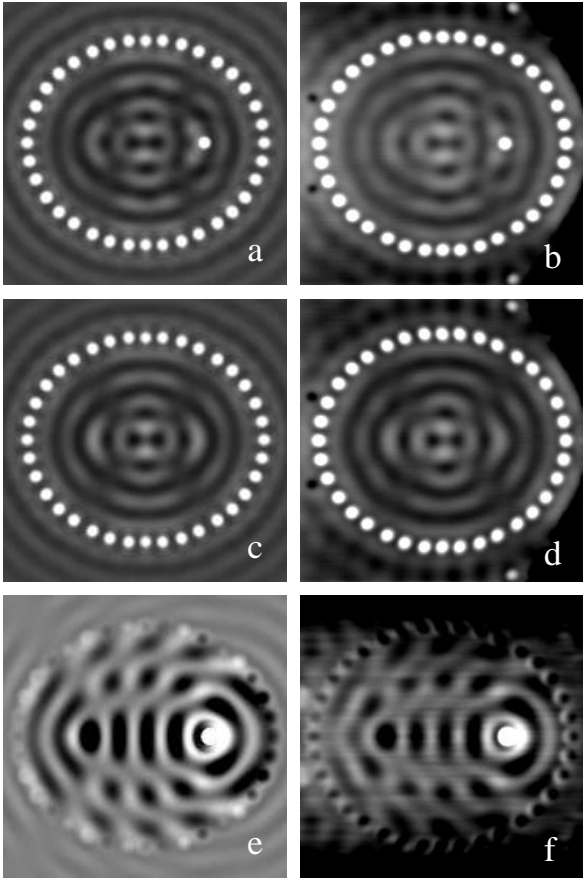}
\caption{Topographical standing wave patterns of a Kondo
corral.  Using the scattering theory and phase shifts described in the
text, these STM topographic images were computed using exact Co adatom
positions on Cu(111) at 4 K.  The agreement between theory ({\bf a},
{\bf c} and {\bf e}) and experiment ({\bf b}, {\bf d} and {\bf f}) is
remarkable.  All the experimental images have been symmetrized by
adding the image to itself after being reflected about its major axis.
Topographic images were calculated by numerically integrating the
LDOS$({\vec r},\epsilon)$ over $\epsilon$ for $E_F \leq \epsilon \leq
E_F + 10$ mV.  This corresponds to the topographic images taken
experimentally in {\bf b} and {\bf d} at a bias voltage of 10 mV. {\bf
e} is the difference of {\bf a} and {\bf c}. {\bf f} is the difference
of {\bf b} and {\bf d}. \label{images_1}}
\end{figure}

\begin{figure}[t]
\epsfxsize=8cm
\epsfbox{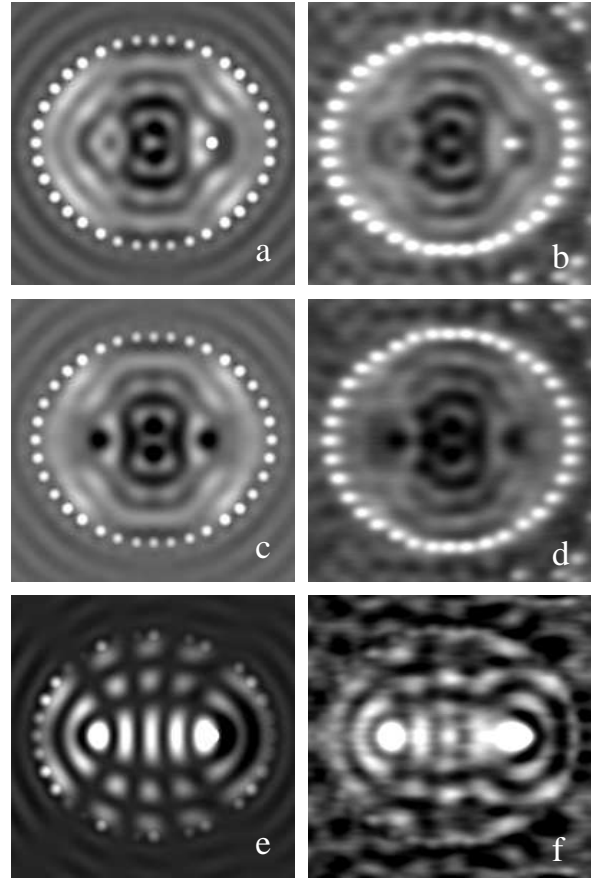}
\caption{dI/dV Standing wave patterns of a Kondo
corral.  Same theoretical vs. experimental arrangement as in
Fig.~\ref{images_1}.  dI/dV measurements were taken simultaneously
with topographic images at a 10 meV bias.  Note that {\bf e} and
{\bf f} resemble an eigenstate of the ellipse.  The ellipse was
constructed to have large surface state amplitudes at the two foci.
\label{images_2}}
\end{figure}

\begin{figure}[t]
\epsfxsize=8cm
\epsfbox{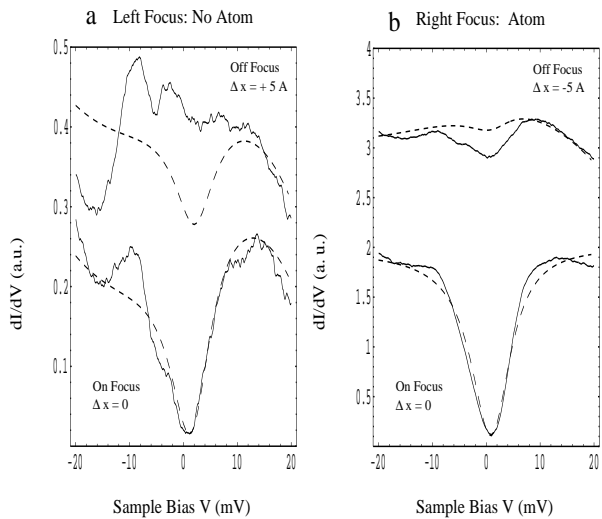}
\caption{Tunneling into the focal atom and empty focus:
The Mirage.  Tunneling spectroscopy is calculated (dashed lines) with
the scattering theory and phase shift given in the text at the empty
focus {\bf a}. Tunneling spectroscopy at the occupied focus is shown
in {\bf b}.  A constant background slope has been removed from both
the experimental data and the calculation.  The attenuation of the
mirage is determined by inelasticity in the scattering of electrons at
the walls of the ellipse.  The theoretical signal 5~\AA\ away from the
empty focus in {\bf a} is lost in the noise of the experiment and is
not a breakdown of the theory.\label{onatom}}
\end{figure}

Applying the scattering theory of Sec.~\ref{sec:scatt} and the phase
shift, Eq.~(\ref{phase}), to elliptical corrals results in the images
shown in Figs.~\ref{images_1} and~\ref{images_2}.
The agreement with 
experiment is excellent.  Our calculation of the tunneling
spectrum at the two foci is compared with experiment in
Fig.~\ref{onatom}.  Note that the signal at the unoccupied focus
is attenuated by approximately a factor of 8, both experimentally and
theoretically.
The calculated spectroscopy in Fig.~\ref{onatom} most clearly
demonstrates that the Kondo mirage is due only to resonant scattering
of electrons from the Co adatom at the opposite focus, even though the
electrons are also resonantly scattering from the wall adatoms:
Calculations performed with $\delta=i\infty$ \cite{rick} instead of
$\delta(\epsilon)$ from Eq.~(\ref{phase}) for the wall atoms show the
Kondo resonances of the wall atoms play no essential role in the
projection of the mirage to the empty focus as the signal in
Fig.~\ref{onatom} is essentially unchanged.  Experimentally the same
result is found when the wall Co adatoms are replaced by
CO~\cite{hari}.
   
Only certain ellipses will give a good mirage effect--those which have
large surface state amplitudes at the foci when the scattering problem
is calculated--and this depends on the relative dimensions of the
ellipse and $\lambda_F$.  Only then will there be appreciable surface
state electron amplitude at the focal adatom to give a strong signal
of Kondo effect in the surface states of Cu(111) at the opposite
focus.\footnote{The relative size of the surface state amplitude at a
given position inside the ellipse also explains why the projection of the
Kondo mirage is insensitive to whether the walls are Kondo (Co) or not
(CO). Near the walls, this amplitude is small in ellipses that have
peak amplitudes at the foci.} Our theory predicts that the quantum
mirage is not restricted to an ellipse or even a ``closed'' structure.
Any time one can construct an arrangement of adatoms or other defects
that lead to a buildup of surface state electron amplitude at two
locations within the coherence length of the electron, a mirage can be
projected.

In conclusion, the quantum mirage reveals no information about local
polarization of the surface state (or bulk) electrons.  The
unpolarized STM cannot measure the size of the Kondo ``screening
cloud'' since it only returns an average signal of spin up and spin
down electrons (or holes) tunneling into the surface.  However, there
are still several important things that can be learned from a
combination of scattering theory and experiment about Kondo impurities
on the surfaces of noble metals.  Firstly, Kondo impurities still act,
to a large extent, like ``black dot'' scatterers.  This is clear from
the appreciable imaginary part of the scattering phase shift given in
Eq.~(\ref{phase}).  The Kondo effect effect does not ``block'' or
inhibit the scattering of surface state electrons into the bulk at the
impurities.\footnote{In principal, the Kondo effect should lessen the
incoherent scattering at the atoms because it tends to ``freeze-out''
the spin, when compared to Fe impurities, for example. (Assuming, of
course, that spin-flip scattering is indeed important at the Fe
impurities.)} Secondly, the Fano line shape of the quantum mirage
can be understood from a resonance in the scattering phase shift with
a non-zero background phase shift.  This complements the ``on atom''
picture of the Fano resonance in dI/dV which can be thought of as
electrons tunneling into both the conduction electron states of the
host (of surface and bulk character) and electrons tunneling into the
``d-level'' of the impurity~\cite{li,madhavan,schiller,plihal}.  For
an STM tip initially above a Kondo impurity, one can think of the Fano
line shape from tunneling as ``rolling over'' to a Fano lineshape from
scattering when the tip moves laterally away from an impurity
\cite{fiete,ujsaghy}.  Thirdly, the fact that the atoms in corral
walls show a Kondo resonance much the same as the resonance from an
isolated impurity on the surface means that the RKKY interactions
between impurities is very weak.  Moreover, the mirage is independent
(both theoretically and experimentally) of the character of the wall
atoms.  In the corrals that show a strong mirage the surface state
electron density is small near the walls, yet the STM signal of the
wall atoms is more or less unchanged.  This suggests that it is mostly
the bulk electrons that are involved in the Kondo effect.  This is the
same conclusion that has been reached recently by~\textcite{knorr} from
studies of a single Kondo impurity.

\section{Related Work and Recent Developments}
\label{sec:recent}
Recently there have been several important developments in the study
of quantum corrals, especially related to the recent mirage
experiments and studies of the lifetimes of quasi-particles in the
surface states.  While our scattering theory explains nearly all of
the observed features of quantum corrals, including the mirage
experiments, it is phenomenological and based on a single-particle
model.  A full understanding of the surface state response to magnetic
impurities requires more detailed studies: Experimentally with spin
resolved STM and theoretically with first principle and many-body
calculations.  It is necessary to go beyond the single-particle
theory, for example, to accurately calculate quantities such as
spin-spin correlation functions of impurities in quantum corrals,
details of the Kondo effect itself or how surface state lifetimes can
be modified by quantum corrals.  Some of these studies have already
been undertaken and we briefly describe them below.

\subsection{Experimental}
Since the mirage experiments, there have been few experimental studies
specific to corrals reported; however,~\textcite{kliewer} have studied
the effect of the modification of surface state electron density by
corrals on the spectroscopy of Mn on Ag(111) and \textcite{kliewer_njp} and
\textcite{braun} have used quantum corrals and related structures to
obtain information about the many-body lifetime effects in the surface
states.  Most STM studies have focused on the Kondo effect from the
impurities themselves. ~\textcite{chen} reported the disappearance of
the Kondo resonance for Co dimers on Au(111). ~\textcite{jamneala}
carried out a systematic study of 3-d elements on Au(111).
~\textcite{teri} reported Kondo effect from Co {\em clusters} adsorbed
on single wall metallic nanotubes.\footnote{The Kondo effect generated
by a ferromagnetic cluster turns out to have several interesting and
nontrivial new features compared to a single impurity \cite{fiete_2}.}
~\textcite{madhavan_2} studied Co on Au(111) as a function of impurity
coverage from isolated impurities up to one
monolayer. ~\textcite{nagoakoa} looked at the temperature dependence
of the broadening of the Kondo resonance of Ti on
Ag(100). ~\textcite{schnieder} measured the scattering phase shift
from isolated Co atoms on Ag(111) and~\textcite{knorr} have studied
the role of surface and bulk state contributions to the Kondo effect
for Co on Cu(100) and Cu(111).

\subsection{Theoretical}
On the theoretical side, much more work has focused on the quantum
mirage in corrals rather than on the single impurities.
~\textcite{agam, porras} and ~\textcite{weissmann} have also developed
theories for the quantum mirage based on a single-particle picture.
More recently, ~\textcite{aligia1} and \textcite{shimada} has
developed a many-body theory of the quantum
mirage. ~\textcite{chiappe} and~\textcite{correa} have undertaken
studies of the interaction between two magnetic impurities in a
quantum corral.  A model of interactions between two impurities in
states confined to the surface of a sphere was studied
by~\textcite{hallberg}.  A recent renormalization group study carried
out by~\textcite{cornaglia} for Kondo impurities in nanoscale systems
also makes contact with the mirage experiments.  A recent work by
\textcite{morr} looks at the quantum mirage from non-Kondo impurities
in a quantum corral built on a superconductor.

While there are now several theories addressing the physics of the
mirage, we feel the least addressed question is that of the relative
role of surface and bulk states in the formation of the Kondo effect
at a single impurity.  Many theories tend to neglect the bulk states
and treat the quantum corral as a confined 2-d system.  We believe
theory should now move beyond this and include the role of both
surface states and bulk states in Kondo resonance.  It remains clear,
however, that the mirage effect is dominated by a Kondo effect that
{\em involves} the surface state electrons because the phase shift,
Eq.~(\ref{phase}), demands it.

\section{Variations of ``Quantum'' Corrals: Optical Corrals and Acoustical Corrals}
\label{sec:variations}
Recently there have been several interesting variations of ``quantum''
corrals.  Most notably, there are now both theory~\cite{girard,wubs} and
experimental realizations~\cite{chicanne} of optical quantum corrals and
related structures.  The theory of optical corrals is quite similar to 
quantum corrals, the main difference being that the electric field is a 
vector field while the wavefunction is a scalar field.  In the optical corrals
the adatoms are replaced by ``posts'' of a different dielectric constant to
confine the electric field.

The same basic physics of quantum corrals also applies to acoustical
corrals in which one can define a LDOS of states that is a local
acoustical impedance function.  The impedance is of course determined by
the same ``in phase'' {\it vs} ``out of phase'' condition of the
returning wave relative to the outgoing wave. A map of acoustical
impedance as a function of position in the room should show exactly
the same type of oscillation with distance as does the STM dI/dV
data. This serves to again remind us that the STM images are not
``snapshots'' of a wave caught in a cavity, like water waves in a
bathtub at some moment.  In fact the analogy of the quantum corrals
and room acoustics is quite close, since a Q-factor of 2 is not unusual
for relatively ``quiet'' rooms.

\section{Conclusions}
\label{sec:conclusions}
In this Colloquium we have reviewed the basic physics of quantum
corrals, including the more recent experiments involving the Kondo
effect.  A single-particle scattering theory with only an s-wave phase
shift is able to account quantitatively for nearly all of the experimental
observations to date, including the quantum mirage.  It is a generic
feature of adatoms on the surfaces of the noble metals that they
strongly couple the surface states to the bulk states.  This appears
in the scattering theory as an imaginary part of the phase shift.
When the adatoms are magnetic and below their Kondo temperature, the
many-body Kondo resonance can be taken into account phenomenologically
with a resonance in the phase shift, Eq.~(\ref{phase}).

The scattering theory that we have presented is valid anywhere more
than $\sim 7$~\AA\ away from an adatom as this is the scale over which
an adatom strongly disturbs the local charge density.  From Kondo
impurities there is Fano resonance in dI/dV that persists as the STM
tip is moved from directly over a Kondo atom to a location 10~\AA\ or
more laterally away from it.  For Kondo impurities, the $\sim 10$~\AA\
spatial extent of the Fano line shape in dI/dV is {\em not a measure
of the Kondo screening cloud}.  To date all reported STM studies of
Kondo impurities have been unpolarized and hence they are insensitive
to local spin polarization.  What the $\sim 10$~\AA\ spatial scale
most likely reflects is the scale over which the STM tip can strongly
couple to the atomic states of the impurities.  Hence, it is a scale
associated with charge rather than spin.  The mirage, therefore,
reflects nothing about local spin correlations at the empty focus of
the elliptical quantum corral.  It is simply a way of probing the
Kondo resonance of the impurity at the opposite focus through coherent
electron propagation in the surface states.  The signal at the empty
focus can be thought of as a ``scattering'' Fano resonance originating
from a resonance piece and a energy independent background piece in
the scattering phase shift of the Kondo atom.
 
The new frontier in quantum corral experiments clearly lies in two
directions: (i) Spin-polarized STM and (ii) Probes of many-body
physics.  There are already several theories that predict strong spin
correlations between impurities in corrals
\cite{chiappe,correa,gyorffy} although none have yet been
experimentally reported.  The relative role of bulk and surface states
in the Kondo effect is still an open question, although experimental
progress has been made~\cite{knorr} which suggest that the Kondo
effect is dominated by bulk states.  Quantum corrals can also provide
tunable environments to study and even modify the physics of the
surface states themselves~\cite{braun,kliewer_njp}. With ever
improving experimental technology we expect to see even more surprises
and fascinating effects to appear in these tiny, engineered
laboratories of many-body physics.

\acknowledgments 
We would especially like to thank our wonderful
experimental collaborators on this subject: M.F. Crommie,
D. M. Eigler, C. P. Lutz and H. C. Manoharan.  Many thanks also go to
I. Affleck, B. I. Halperin, J. S. Hersch, S. Kehrein, Y. Oreg,
O. \'Ujs\'aghy, G. Zar\'and and A. Zawadowski. This work was supported
by NSF grants PHY-0117795 and CHE-0073544.

\bibliography{fiete}

\end{document}